\documentstyle[editedvolume,epsfig]{crckapb}

\begin{opening}
\title{TIMING THE KILOHERTZ QUASI-PERIODIC\protect\\
OSCILLATIONS IN LOW-MASS X-RAY BINARIES}

\author{Mariano M\'{e}ndez}
\institute{Astronomical Institute ``Anton Pannekoek'',\\
       University of Amsterdam,\\
       Kruislaan 403, NL-1098 SJ Amsterdam, the Netherlands}

\end{opening}

\runningtitle{TIMING THE KHZ QPOS IN LMXBS}

\begin{document}

\section{Introduction}

In the past 3 years the Rossi X-ray Timing Explorer has discovered
kilohertz quasi-periodic oscillations (kHz QPOs) in the persistent flux
of 19 low-mass X-ray binaries (LMXBs).  The power density spectra of
most of these sources show twin kHz peaks.  Initial results seemed to
indicate that the frequency separation between the twin kHz peaks,
$\Delta \nu = \nu_{2} - \nu_{1}$, remained constant even as the peaks
gradually moved up and down in frequency (typically over a range of
several hundred Hz) as a function of time.  But thanks to precise
measurements of the frequencies of the kHz QPOs, we know now that in
several sources $\Delta \nu$ is not constant, but it decreases as
$\nu_{1}$ and $\nu_{2}$ increase.

In this chapter I describe a new technique that we have been using in
the past few years to get precise measurements of the frequency
separation of the kHz QPOs in some LMXBs.  My plan is to show how this
technique (that we call ``shift-and-add'') works, and to present some of
the results we obtained using it.  It is not my purpose, however, to
discuss here the details of the kHz QPO phenomenon, as this subject is
extensively covered elsewhere in this book (see the chapter by van der
Klis).

\section{The Average Power Spectrum}

The standard Fourier techniques in X-ray timing are described in detail
in van der Klis (1989a); let us just recall here the definition of the
power spectrum.  A time series $x(t)$ is first divided into $N$ time
intervals of length $\tau$, and for each interval the Fourier transform
is calculated as:

\begin{equation}
X_{n}(\nu) = \int_{t_{0}+n \tau}^{t_{0}+(n+1) \tau} x(t) e^{2 \pi i \nu
t} dt, n=0,1,2,...,N,
\end{equation}
where $i = \sqrt{-1}$, and $t_{0}$ is some initial time.  The power
spectrum of $x(t)$ is defined as the average of the square of the
Fourier transforms, $P(\nu) = <X^{2}_{n}(\nu)>$.  The average is
computed to improve the signal-to-noise ($S/N$) ratio of the power
spectrum (see van der Klis 1989a for details).  This definition assumes
that the time series $x(t)$ is {\it stationary}, i.e.  that {\it all}
time segment of length $\tau$ have the same statistical properties.

\begin{figure}
\centerline{\epsfig{figure=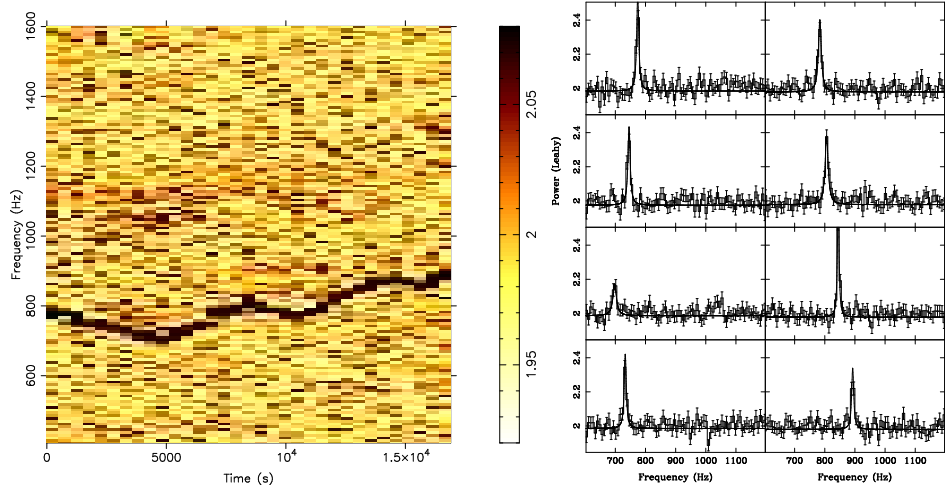,clip=}}
\caption{{\it Left Panel}:  Dynamic power spectrum of the source
4U\,1728--34.  {\it Right Panel}:  Power spectra of some selected
intervals.}
\end{figure}

As I already mentioned, one property of the kHz QPOs is that their
frequencies vary with time (therefore the underlying time series is
non-stationary).  This is shown in Figure 1, for the source
4U\,1728--34.  On the left panel I show a dynamical power spectrum,
where the x-axis represents time (the origin of time in this case is
1997 October 1 at 06:09 UTC), the y-axis is the Fourier frequency, and
the different gray levels indicate the power in each frequency bin.  The
dark feature that crosses the Figure almost horizontally is one of the
two QPOs detected in this source.  The right panel shows power spectra
($P(\nu)$ vs.  Fourier frequency) at some selected times:  The same QPO
is seen moving between $\sim 700$ and 900 Hz.

\begin{figure}
\centerline{\epsfig{figure=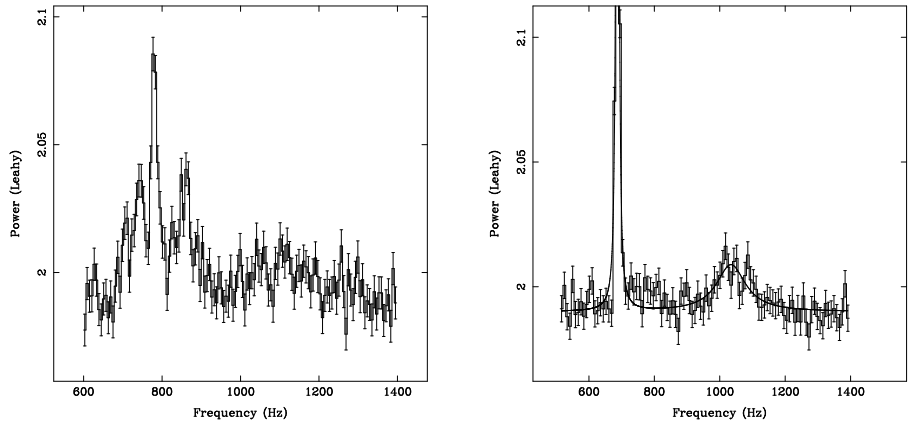,width=12cm,clip=}}
\caption{{\it Left Panel}:  Average (standard) power spectrum of the
same data shown in Figure 1.  {\it Right Panel}:  Shifted-and-averaged
power spectrum of the same data.}
\end{figure}

Obviously, the average power spectrum of the whole observation will show
several peaks, at the frequencies at which the QPO stays for a while
during the observation.  This is shown in Figure 2 left panel, where the
average power spectrum has been calculated as described above.  However,
as the QPO is strong enough to allow us to measure it in short time
segments, we can think of ``shifting'' the frequency axis of each
individual power spectrum before we calculate the average so that the
new power spectrum is $\tilde P(\nu) = <X^{2}_{n}(\nu - \nu_{0,n})>$.
Here $\nu_{0,n}$ is the frequency shift that we need to apply to the
power spectrum of each segment if we want to align the strong QPO at the
same frequency before we calculate the average.  Notice that this is
equivalent to multiplying the function $x(t)$ by $e^{-2 \pi i \nu_{0,n}
t}$ before calculating the integral in eq.(1).

The right panel in Figure 2 shows the result of such procedure (which,
of course, also implies measuring $\nu_{0,n}$ for each segment;
certainly this is extra work, but it comes with some rewards, as I will
explain in \S3).  If it were only for the strong QPO in the average
power spectrum (this QPO is stronger in the new power spectrum than in
the original one) we would have not gained much by this procedure.
Although there is now only one sharp peak, this shifted-and-averaged
power spectrum provides no new information about the strong QPO.
However, there is a second (weaker, but yet significant) QPO in the
shifted power spectrum, $\sim 350$ Hz above the strong QPO.  Although
there seems to be evidence for such a weak QPO in the original power
spectrum (Fig.  2, left panel), there it was not significant enough.

To understand why we detect this QPO in the shifted power spectrum,
while in the standard power spectrum it was not significant enough, we
need to recall that the $S/N$-ratio of the QPO is proportional to
$(T/W)^{1/2}$, where $T = N \times \tau$ is the total length of the
observation, and $W$ is the width of the QPO (see van der Klis 1989b).
If the frequency separation between the 2 QPOs is roughly
constant\footnote{Although it is not really constant, variations in
$\Delta \nu$ are smaller than changes in the centroid frequencies of the
QPOs.}  (a close inspection to Fig.  1 left panel shows that there might
be such weak QPO in the dynamic power spectrum, ``following'' the strong
QPO as this one moves in frequency), the shift applied to the power
spectra of the individual segments to align the strong QPO also aligns
the weak QPO.  This makes this new QPO narrower, and therefore more
significant, in $\tilde P(\nu)$ than in $P(\nu)$.

This is how we first discovered the second QPO during 1996 outburst of
4U\,1608--52 (M\'{e}ndez et al.  1998a), which we afterwards detected
again (this time without requiring the sensitivity-enhancing technique
described above) during another outburst in 1998 (M\'{e}ndez et al.
1998b).

Not surprisingly, this same technique can be used to obtain much better
measurements of the frequency separation between the kHz QPOs.  This is
because the error in the centroid frequency of a QPO is proportional to
the width of the QPO, and inversely proportional to its S/N-ratio (see
Downs \& Reichley 1983 for a similar situation when measuring the
arrival time of individual pulses in radio pulsars).  As described
above, the weak QPO is usually narrower in the shifted power spectrum,
$\tilde P(\nu)$, than in the standard power spectrum, $P(\nu)$, and
therefore its frequency can be measured more accurately in $\tilde
P(\nu)$ than in $P(\nu)$.

One way of measuring $\Delta \nu$ using the shifted power spectrum is
the following:  Divide the data into $N$ segments of length $\tau$, and
for each segment calculate the power spectrum.  Measure the frequency of
the strong QPO, $\nu_{1}$, in each of these $N$ power spectra, and group
the data in $M$ sets such that $\nu_{1}$ is more or less constant within
each set.  This ensures that $\Delta \nu$ is more or less constant
within each set\footnote{Notice that there is a trade-off here:  the
choice of small frequency intervals for $\nu_{1}$ ensures small changes
in $\Delta \nu$, preventing artificial broadening of the weak QPO when
the power spectra are shifted; but it also reduces the amount of data
that are averaged in each set, and therefore decreases the $S/N$-ratio
of the weak QPO at $\nu_{2}$.}  (early results in Sco\,X--1 showed that
$\Delta \nu$ decreased monotonically as $\nu_{1},\nu_{2}$ increased; van
der Klis et al.  1997).  For each set, align the power spectra of the
individual time segments using $\nu_{1}$ as a reference, and combine
them to produce an average power spectrum.  The result is $M$ (aligned)
power spectra for which both kHz QPOs are as narrow as possible.  For
each of these $M$ power spectra measure $\nu_{1}$ and $\nu_{2}$,
calculate $\Delta \nu = \nu_{2} - \nu_{1}$, and plot $\Delta \nu$ vs.
the average value of $\nu_{1}$ in each set.

\begin{figure}
\centerline{\epsfig{figure=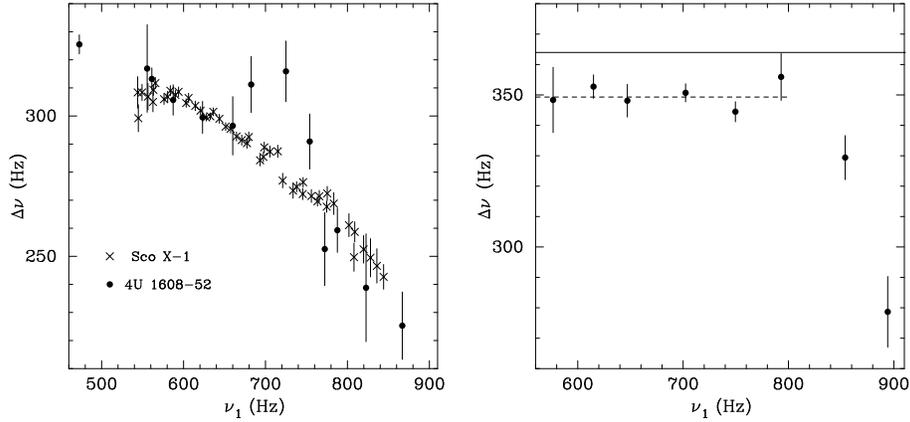,angle=0,width=12cm,clip=}}
\caption{{\it Left Panel}:  Frequency separation between the two
simultaneous kHz QPOs in 4U\,1608--52 (circles) and Sco\,X--1 (crosses).
{\it Right Panel}:  Frequency separation between the two simultaneous
kHz QPOs in 4U\,1728--34 (circles).  The solid line indicates the
frequency of the burst oscillations at 363.95 Hz (Strohmayer et al.
1996).  The dashed line indicates the average value of $\Delta \nu$
(349.3 Hz) for the first 6 points.  For the three sources $\Delta \nu$
as a function of $\nu_{1}$ was calculated as explained in the text.}
\end{figure}

Figure 3 shows the results of applying such procedure to the kHz QPOs in
4U\,1608--52, 4U\,1728--34, and Sco\,X--1.  The implications of these
results (specially in the case of 4U\,1728--34) to the existing models
for the kHz QPOs are presented elsewhere in this book (see the chapter
by van der Klis), and will not be discussed here.

\section{Other Results}

As I mentioned in \S2, there is a reward for the extra work of measuring
$\nu_{1}$ in the individual time segments:  We can combine these
measurements with, for instance, the source count rate or X-ray colors
in those same segments.  One example of this is shown in Figure 4.  The
left panel shows a plot of $\nu_{1}$ vs.  the $2-16$ keV count rate for
4U\,1608-52.  Each point represents 128 s of data.  For the same data
used in the plot on the left panel, the right panel shows the relation
between $\nu_{1}$ and an X-ray color (see the caption of this figure for
the definition of the color).

\begin{figure}
\centerline{\epsfig{figure=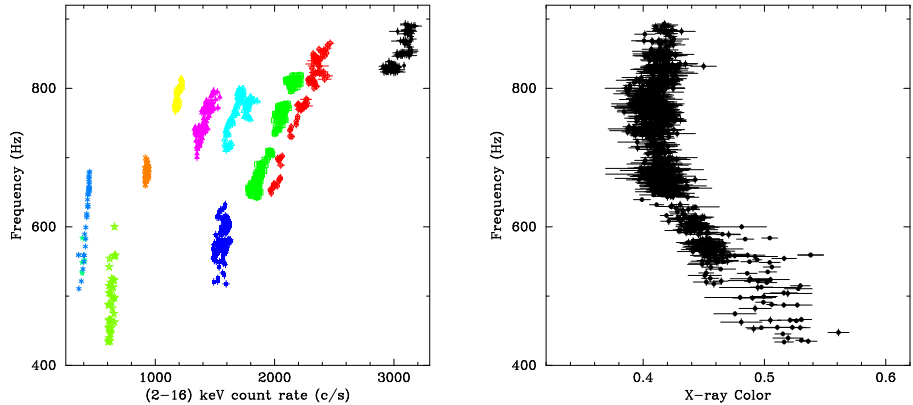,angle=0,width=12cm,clip=}}
\caption{{\it Left Panel}:  Relation between $\nu_{1}$ and the $2-16$
keV count rate in 4U\,1608--52.  {\it Right Panel}:  For the same data
as on the left panel, $\nu_{1}$ vs.  X-ray hard color, defined as the
ratio of the count rate in the $6.0 - 9.7$ keV band to the count rate in
the $9.7 - 16.0$ keV band.}
\end{figure}

Interesting in this Figure is the complex dependence of $\nu_{1}$ upon
X-ray count rate, which is usually assumed to be a good measure of mass
accretion rate $\dot M$, and how this complexity is reduced to a single
track in the frequency vs.  hard color diagram.  As before, the
interested reader can find an extensive discussion of this subject
elsewhere in this book.  I just want to say here that these are ``side''
results of the technique described in \S2.

\section{Conclusion}

I presented a new technique aimed at increasing the sensitivity to weak
moving QPOs in the power density spectra of LMXBs.  I showed some
examples of the results obtained by applying this technique to data
obtained with the Rossi X-ray Timing Explorer.  In particular, this
technique provides more precise measurements of the frequencies of the
kHz QPOs, which can be used to set constraints on the models so far
proposed to explain this new phenomenon.

%\vskip0.1cm
%\noindent
\acknowledgements{
%{\it Acknowledgments}
This work was supported by the Netherlands
Research School for Astronomy (NOVA), the Netherlands Organization for
Scientific Research (NWO) under contract number 614-51-002, the Leids
Kerkhoven-Bosscha Fonds (LKBF), and the NWO Spinoza grant 08-0 to E.P.J.
van den Heuvel.  MM is a fellow of the Consejo Nacional de
Investigaciones Cient\'{\i}ficas y T\'{e}cnicas de la Rep\'{u}blica
Argentina.  This research has made use of data obtained through the High
Energy Astrophysics Science Archive Research Center Online Service,
provided by the NASA/Goddard Space Flight Center.}

\end{document}